\documentclass{aip-cp}

\usepackage[numbers]{natbib}
\usepackage{rotating}
\usepackage{graphicx}

\begin{document}

\title{The influence of intruder states in even-even Po isotopes}

\author[aff1]{J.E. Garc\'{\i}a-Ramos
}
\eaddress{enrique.ramos@dfaie.uhu.es}
\author[aff2]{K.~Heyde}
\eaddress{kris.heyde@ugent.be}

\affil[aff1]{Departamento de F\'{\i}sica Aplicada, Universidad de
    Huelva, 21071 Huelva, Spain}
\affil[aff2]{Department of Physics and Astronomy, Ghent University,
    Proeftuinstraat 86, B-9000 Gent, Belgium}

\maketitle

\begin{abstract}
We study the role of intruder states and shape coexistence in the
even-even $^{190-206}$Po isotopes, through an interacting
boson model with configuration mixing calculation. We
analyzed the results in the light of known systematics on various observable in the 
Pb region, paying special
attention to the unperturbed energy systematics and quadrupole deformation. We
find that shape coexistence in the Po isotopes behaves in very much the same way 
as in the Pt isotopes, i.e., it is somehow hidden, contrary to the situation in the Pb 
and the Hg isotopes. 
\end{abstract}

\section{INTRODUCTION}

Shape coexistence has been observed in many mass regions
throughout the nuclear chart and has become a very useful paradigm
to explain the competition between the monopole part of the nuclear effective force that tends to
stabilize the
nucleus into a spherical shape, in and near to shell closures, and the strong correlations (pairing,
quadrupole in particular) that favors the nucleus into a deformed shapes in around mid-shell regions
\cite{heyde11}.   

New experimental results pushed our knowledge on the
even-even Pt, Hg, Pb, and Po isotopes towards region far from the region of stable nuclei, 
for which experimental information was lacking and highly needed.  
In this mass region the intruder bands are easily singled out for the Pb and Hg isotopes 
and the excitation energies display the characteristic parabolic pattern with
minimal excitation energy around the N$=104$ neutron mid-shell
nucleus. However, this structure seems absent in the Pt and Po isotopes. 
Indeed, in the case of Pt, instead of exhibiting a parabolic shape in the
energy systematics, there appears
a sudden drop in the excitation energies for certain states: 0$^+_2$, 4$^+_1$, 
 2$^+_3$ and 6$^+_1$, followed by a flat behavior around midshell.
In the case of Po, it is even more difficult to disentangle the interplay
between regular and intruder states because the detailed spectroscopic data is not available 
going below neutron number N=106 (A=190), i.e., the mid-shell nucleus is not reached
experimentally. In particular, the energy of the $2_1^+$ state remains 
rather constant ($\sim 600-650$ keV) in the mass range A=196-208, with a
sudden drop at A=196, with corresponding energy of $\sim 300$ keV.  

In a series of articles, we studied the Pt
isotopes extensively \cite{Garc09}, 
as well as the Hg isotopes \cite{Garc14},
with the Interacting Boson Model (IBM) \cite{Iach87} incorporating
proton 2p--2h excitations (IBM-CM) \cite{duval82}. The conclusion of
these studies was that   
configuration mixing in the Pt isotopes is somehow ``concealed'', while
in the case of Hg isotopes (as well as for Pb nuclei) the presence of
intruder states is self-evident. 

The IBM-CM allows the simultaneous treatment and mixing of several
boson configurations corresponding to different particle--hole
shell-model excitations \cite{duval82}. 
Hence, the model space  corresponds to a $[N]\oplus[N+2]$ 
boson space. The boson number $N$ is obtained as the sum of the number
of active protons (counting both proton particles and holes) and the
number of valence neutrons, divided by two. 
Thus, the Hamiltonian for
two-configuration mixing is written
\begin{equation}
  \hat{H}=\hat{P}^{\dag}_{N}\hat{H}^N\hat{P}_{N}+
  \hat{P}^{\dag}_{N+2}\left(\hat{H}^{N+2}+
    \Delta^{N+2}\right)\hat{P}_{N+2}\
  +\hat{V}_{\rm mix}^{N,N+2}~,
\label{eq:ibmhamiltonian}
\end{equation}
where $\hat{P}_{N}$ and $\hat{P}_{N+2}$ are projection operators, 
$\hat{V}_{\rm mix}^{N,N+2}$  describes
the mixing both subspaces,
$\hat{H}^i$ is the IBM Hamiltonian
with $i=N,N+2$, 
and $\Delta^{N+2}$ is related
to the energy needed to excite two particles across the
shell gap.

Within this formalism we have performed a fit to the excitation
energies and $B(E2)$ transition rates of $^{190-208}$Po in order to
determine the parameters for the IBM-CM Hamiltonian. 
The results from the fitting procedure
are summarized in Table II of Ref.~\cite{Garc15b} and are used
in the calculations and results discussed in the present contribution. 
Here, we present IBM-CM calculations for the Po
isotopes, focusing on excitation energies as well as on the quadrupole deformation,
and, compare the corresponding results in the Pt and Hg isotopic series.  

\section{ABSOLUTE ENERGIES AND ENERGY SYSTEMATICS}
\label{sec-energies}
Intruder states are expected to appear, in principle, at an excitation
energy well above the energy of the corresponding regular ones. Therefore, at a first glance, 
the ground state would
always be of regular nature. The reason is that intruder configurations are related
to the creation of 2p-2h excitations across the Z=82 closed
shell. These energies correspond to $\Delta^{N+2}=3480$ keV \cite{Garc14},
$2400$ keV \cite{Garc15b}, and $2800$ keV \cite{Garc09}, for the Hg, Po, and Pt isotopes,
respectively. However, this energy gap has to be corrected by the 
pairing and quadrupole energy gain that can reduce notably this energy difference.  
In particular, around the mid-shell point at N=104 where the number of
active nucleons becomes maximal, the energy gain due to the
strong correlation energy is such that the energy of the intruder
configurations   
becomes close, or even crosses the energy of the regular ones. 
On top of this, there is the mixing between both configurations, that can become
large, making it difficult, in some cases, to
determine which configurations are dominant in the ground-state
wavefunction. 
\begin{figure}[h]
  \centerline{\includegraphics[width=.42\textwidth]{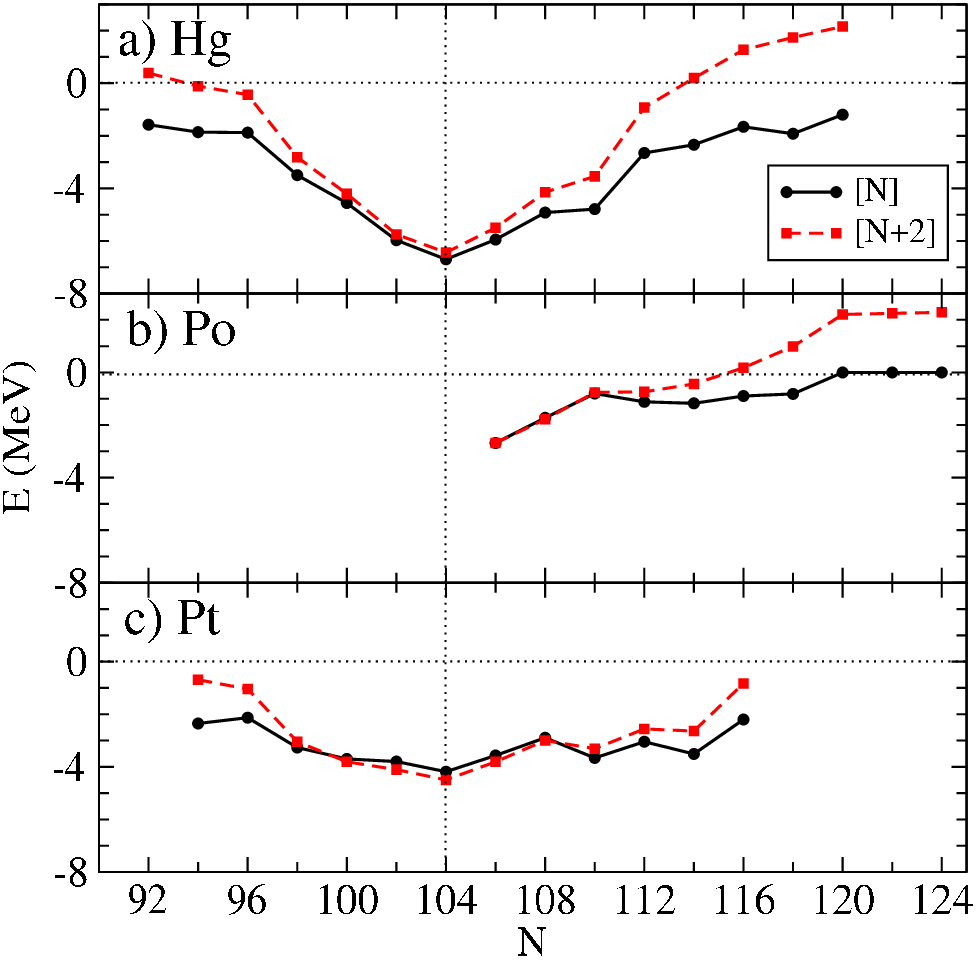}}
  \caption{Absolute energy of the lowest regular [N] and intruder
    states [N+2]
    for $^{172-200}$Hg (panel a), $^{190-208}$Po (panel b), and
    $^{172-194}$Pt (panel c). The same energy scales are used for the three panels.}
\label{fig-corr}
\end{figure}

To face this problem, we calculate
the ``absolute'' energy of the lowest $0^+$ state belonging
to both the [N] regular and [N+2] intruder configuration spaces, turning off the
interaction amongst the two families ($\hat{V}_{\rm mix}^{N,N+2}=0$). 
The energy of the lowest 0$^+$ state in the regular configuration
space [N] and in the intruder configuration space [N+2] 
is lowered by their specific correlation
energy. In general the energy reduction is larger for the intruder configuration as
compared with the regular configuration, mainly because the number of
bosons is larger. 
Therefore, the nature 
of the ground state will depend on the detailed balance between the value of
$\Delta^{N+2}$ and the correlation energies. In Fig.\ref{fig-corr} we
compare the unperturbed (with no mixing) first regular and intruder
$0^+$ states, for Hg (panel a), Po (panel b), and Pt (panel c). First,
one notices that for the Hg isotopes, the lowest configuration is always the
regular one, even at midshell. On the other hand, for Pt isotopes, in most 
cases the regular configuration describes the ground state, except around
midshell, where the intruder configuration takes over to describe the ground state. 
Finally, one 
appreciates for the Po isotopes, a strong similarity with the Pt isotopes, with the
intruder configuration becoming the ground state around midshell. There is a difference though, 
because in the Po isotopes both configurations are almost degenerate in energy  while in the
Pt isotopes, the intruder configurations appear well below the regular configurations around midshell. 
A full calculation with the analysis of the wave functions confirms the above
conclusions (see Refs.~\cite{Garc09,Garc14,Garc15b}). 

\begin{figure}
  \centerline{\includegraphics[width=.5\textwidth]{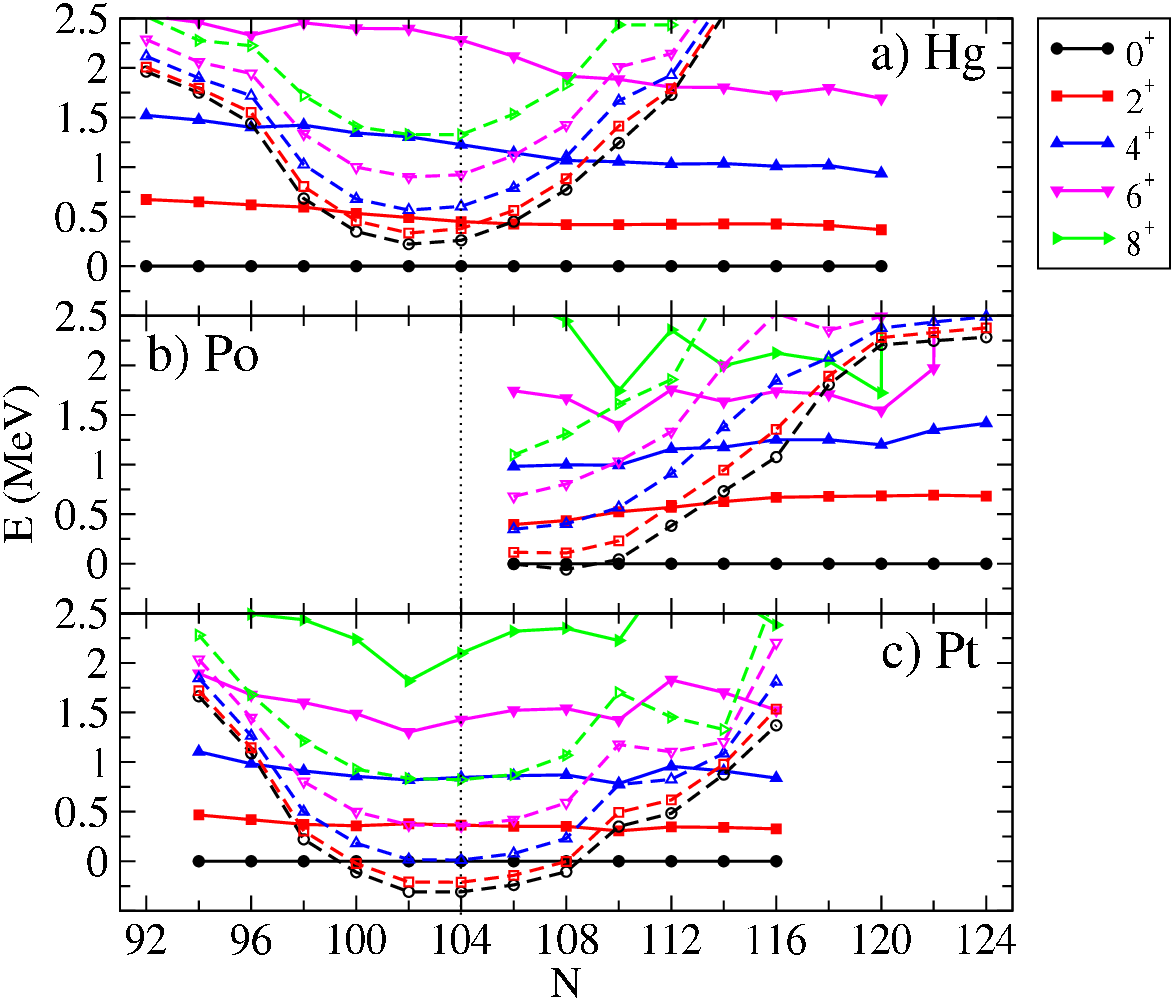}}
  \caption{Comparison of the energy spectra for the IBM-CM Hamiltonian,
    switching off the mixing term. The two lowest-lying regular states and
    the lowest-lying intruder state for each of the angular momenta are
    shown (full lines with closed symbols for the regular states while
    dashed lines with open symbols are used for the intruder ones). 
  }
\label{fig-spectra}
\end{figure}

Regarding the experimental energy systematics of the intruder states, one 
expects a parabolic shape centered around N=104. However, this shape
can be strongly perturbed because of the mixing between regular and
intruder configurations and because of the crossing in energy of these
two configurations, implying a change in the ground-state wavefunction. 
To obtain better insight into the energy systematics of
intruder and regular states, it is 
enlightening to calculate the energy spectra as a function of N
switching off the mixing part of the Hamiltonian. In
Fig.~\ref{fig-spectra} we depict the spectra, separating regular and
intruder states, for the Hg (panel a), Po (panel b), and Pt (panel
c) isotopes. Comparing the three isotopic series, one can appreciate a common picture 
that cannot be observed easily inspecting the experimental energy 
spectra (except in the case of Hg due to its small mixing term). 
In fact, in the three isotope series, the intruder configurations indeed 
present the expected 
parabolic shape, while the regular ones exhibit a rather flat energy
behavior. The 
only, but important, difference between the three panels is the relative position 
in energy of the intruder and regular states. 
Note that for the Po and Pt isotopes, it is not
possible to clearly see, from the experimental data solely, a different
pattern in the energy systematics of regular and intruders
because of the strong mixing and due to the
crossing of regular and intruder configurations in the ground state
around midshell.  

\section{NUCLEAR DEFORMATION}
\label{ßec-deformations}
In this section, we use a phenomenological approach    
to extract information on the quadrupole deformation $\beta$ for both
the regular and the intruder configurations. To this end we rely
on the geometrical picture of the nucleus in order to extract the $\beta$ parameter
for a given band, either regular or intruder, from known $B(E2)$ values
\begin{equation}
\beta=\frac{4\pi\sqrt{B(E2: I_i\rightarrow I_f)}}{3 \langle I_i K 2 0| I_f
  0\rangle Z e R_0^2},
\end{equation} 
where $R_0=1.2 A^{1/3}$ fm and $<.... | ..>$ is the Clebsch-Gordan
coefficient. 
The key point is the assumption that around midshell, the $0_1^+$ and
$2_1^+$ states belong to the regular (or intruder) band while the $6_1^+$ and
$4_1^+$ states belong to the
intruder (or regular) band. Therefore, using the corresponding E2 transitions, one can
extract in a separate way the deformation corresponding with the intruder and regular
configurations. When removed far from midshell, most probably both
transitions will involve regular states because in this region
intruder are supposed to be well above in energy (see also Fig.~\ref{fig-spectra}).

\begin{figure}
  \centerline{\includegraphics[width=.42\textwidth]{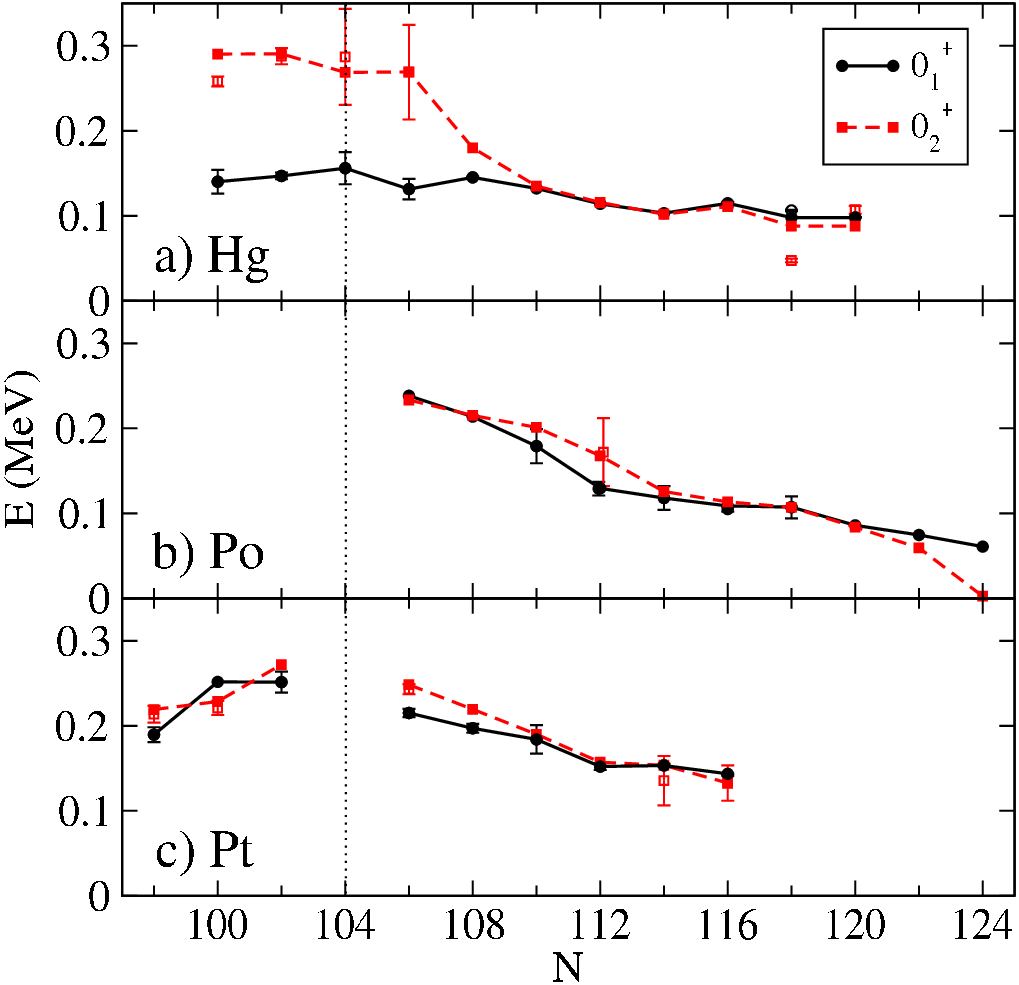}}
  \caption{Comparison of the value of $\beta$ extracted from
    experimental and theoretical B(E2) values corresponding to the
    $0_1^+$ and $0_2^+$ states. 
}
\label{fig-beta}
\end{figure}

In Fig.~\ref{fig-beta}, we depict the value of
$\beta$ using the above method for the Hg (panel a), the Po (panel b), and the Pt
(panel c) isotopes. In the case of Hg, one easily singles out the presence of
two configurations with very different deformation; the lower
configuration, which corresponds to the regular state, is less deformed
than the higher one and can be identified with an intruder
state. This is an empirical evidence about the nature of the
ground state which, close to midshell, always corresponds to
a less deformed and regular configuration. In the Po and Pt isotopes, 
a completely different systematics compared with Hg shows up. Both
configurations present a rather similar deformation, which is a
consequence of the large mixing between both families of
configurations and therefore it is not possible to discriminate
them. The maximum deformation is reached around midshell where a bump
is observed. 

\section{CONCLUSIONS}
In this contribution, we have, starting from new IBM-CM calculations 
performed for the Po isotopes \cite{Garc15b}, compared those results 
with previous results obtained for the Hg and Pt isotopes using the same
approach. Our main conclusion is the strong 
similarity between the Po and Pt isotopes as well as striking differences with 
the Hg isotopes (as well as with the Pb isotopes). In the Po
isotopes, strong mixing between the regular and intruder
configurations results from the calculations, with the intruder configuration becoming 
dominant in describing the 0$^+$  ground state wavefunction 
around midshell. In summary, configuration mixing is concealed in the
Po isotopes, very much like it is the case for the Pt isotopes, 
while in Hg, the presence of the intruder configurations has become self-evident.  

\section{ACKNOWLEDGMENT} 
Financial support from the ``FWO-Vlaanderen'' (KH and JEGR) and the
InterUniversity Attraction Poles Programme - Belgian State - Federal
Office for Scientific, Technical and Cultural Affairs (IAP Grant No.
P7/12, is acknowledged.  This work has also been partially supported
by the Spanish MINECO and the FEDER under Project No.\
FIS2011-28738-C02-02 and by Spanish Consolider-Ingenio 2010 
(CPANCSD2007-00042).


\nocite{*}
\bibliographystyle{aipnum-cp}%

\end{document}